\documentclass[aps,prd,twocolumn, showpacs,
  superscriptaddress]{revtex4}

\usepackage{graphicx} \usepackage{color} \usepackage{natbib}
\usepackage{epsfig}
\usepackage{amsmath,amssymb,amsfonts}
\def\prl{Phys. Rev. Lett.}
\def\prd{Phys. Rev. D}
\def\cqg{Class. Quantum Grav.}

\def\apj{Astrophys. J.}

\def\mnras{Mon. Not. R. Astron. Soc.}

\begin{document}

\title{Schwarzschild-de Sitter Spacetimes, McVittie Coordinates, and Trumpet Geometries}

\author{Kenneth A. Dennison}
\author{Thomas W. Baumgarte}
\affiliation{Department of Physics and Astronomy, Bowdoin College, Brunswick, ME 04011, USA}

\begin{abstract}
Trumpet geometries play an important role in numerical simulations of black hole spacetimes, which are usually performed under the assumption of asymptotic flatness.  Our Universe is not asymptotically flat, however, which has motivated numerical studies of black holes in asymptotically de Sitter spacetimes.  We derive analytical expressions for trumpet geometries in Schwarzschild-de Sitter spacetimes by first generalizing the static maximal trumpet slicing of the Schwarzschild spacetime to static constant mean curvature trumpet slicings of Schwarzschild-de Sitter spacetimes.  We then switch to a comoving isotropic radial coordinate which results in a coordinate system analogous to McVittie coordinates.  At large distances from the black hole the resulting metric asymptotes to a Friedmann-Lema\^itre-Robertson-Walker metric with an exponentially-expanding scale factor.  While McVittie coordinates have another asymptotically de Sitter end as the radial coordinate goes to zero, so that they generalize the notion of a  ``wormhole'' geometry, our new coordinates approach a horizon-penetrating trumpet geometry in the same limit.  Our analytical expressions clarify the role of time-dependence, boundary conditions and coordinate conditions for trumpet slices in a cosmological context, and provide a useful test for black hole simulations in asymptotically de Sitter spacetimes. 
\end{abstract}


\maketitle

\section{Introduction}
\label{Intro}

Numerical simulations of binary black hole mergers, without any symmetry assumptions, became possible with the pioneering work of \cite{Pre05b,CamLMZ06,BakCCKM06a}.  One particularly simple approach to handling the spacetime singularities at the centers of the black holes is to adopt so-called moving-puncture coordinates in the context of a spatial foliation of the spacetime (see, e.g. \cite{CamLMZ06,BakCCKM06a} and numerous later references; see also \cite{BauS10} for a textbook treatment.)  Essentially, these coordinates avoid the spacetime singularities by driving the spatial slices toward what is now referred to as a trumpet geometry (see, e.g., \cite{HanHPBO06,HanHOBO08,Bro08}).

Trumpet geometries have very special properties.  In the Schwarzschild spacetime \cite{Sch16}, trumpet slices end on a limiting surface of finite areal radius, say $R_0$, and hence do not reach the spacetime singularity at $R = 0$.  Simultaneously, any point away from the limiting surface, with $R > R_0$, is an infinite proper-distance away from the limiting surface (see \cite{DenBM14} for a characterization of trumpet geometries in Kerr spacetimes \cite{Ker63}).  In an embedding diagram (see Fig.~2 of \cite{HanHOBO08} for an example) the shape of these surfaces resembles a trumpet, which explains their name.

Trumpet geometries are not unique, and can have different mean curvatures.  Most common in numerical simulations are trumpets in a ``stationary 1+log'' slicing (e.g. \cite{HanHPBO06,HanHOBO08,Bru09}), while maximally sliced trumpet geometries \cite{HanHOBGS06} are easier to analyze analytically \cite{BauN07}.  More recently,  we also found a remarkably simple family of analytical trumpet slices of the Schwarzschild spacetime \cite{DenB14}, which can even be generalized for rotating Kerr black holes \cite{DenBM14}.

Most work on trumpet geometries applies to asymptotically flat spacetimes.  Our Universe is not asymptotically flat, however, and numerous numerical relativity simulations have been performed in a cosmological context
\cite{NieJ99,ShiS99,MusM13,RekCF15a,RekFC16,GibMS16,BenB16}). In particular, cosmological observations suggest that we live in a Universe that is dominated by dark energy, or a cosmological constant with a small positive value (see, e.g., \cite{RieFCCDGGHJKLPRSSSSSST98, PerAGKNCDFGGHKKLNPPQ99} and numerous later references), so that, in the future, the large-scale structure of our Universe will be increasingly well approximated by that of a de Sitter (dS) spacetime \cite{DeSit17a,DeSit17b}.  This, in turn, has motivated numerical relativity simulations in asymptotically dS spacetimes, including the binary black hole merger simulations of  \cite{ZilCGHSW12}.

For single black holes, the authors of \cite{ZilCGHSW12} adopt as initial data a Schwarzschild-de Sitter (SdS) solution \cite{Kot18,Wey19,Tre22} expressed in so-called McVittie coordinates \cite{McV33} (see also, e.g. \cite{KalKM10,LakA11} for more recent discussions) which generalize the ``wormhole" data familiar from Schwarzschild spacetimes to SdS spacetimes (see Section \ref{sec:mcvittie} below).   They also adopt moving-puncture coordinates, so that one would expect dynamical simulations to settle down to a SdS generalization of a trumpet geometry.  This raises the question whether this final, asymptotic state of black hole simulations can be expressed analytically, as in the asymptotically flat case.

In this paper we derive and explore constant mean curvature trumpet geometries in SdS spacetimes and generalize the results of \cite{BauN07} to SdS spacetimes.  We introduce McVittie-like coordinates which may be better suited for numerical simulations in some respects, but also introduce a time-dependence.   We demonstrate how our analytical solution can be used as a test of numerical black hole simulations in asymptotically dS spacetimes, and hope that these results will also be helpful in the interpretation of future numerical simulations.

This paper is organized as follows.   In Sec.~\ref{SdS} we present a sequence of coordinate transformations for SdS spacetimes, starting with static coordinates in Sec.~\ref{sec:static} and McVittie coordinates in Sec.~\ref{sec:mcvittie}, and finally ending with trumpet slices expressed in McVittie-like coordinates in Sec.~\ref{sec:mcvittietrumpet}.  In Sec.~\ref{sec:numerics} we show results from a numerical implementation, showing how our analytical results can be used as a testbed for black hole simulations in asymptotically dS spacetimes.  We briefly summarize in Sec.~\ref{Sum}.   Throughout this paper we adopt geometric coordinates with $c = G = 1$.

\section{Slicing Schwarzschild-de Sitter Spacetimes}
\label{SdS}

\subsection{Static coordinates}
\label{sec:static}

The line element for  SdS spacetimes is often given in static coordinates as
\begin{equation}
\label{static}
ds^{2} = -f dT^{2}+ f^{-1}dR^{2} + R^{2}d\Omega^{2},
\end{equation}
where we have abbreviated $f\equiv1-2M/R-H^{2}R^{2}$ (see, e.g., \cite{SprSV01} for an introduction to dS and SdS spacetimes in various coordinate systems).  Here $M$ is the black hole mass, $R$ the areal radius, and the Hubble parameter $H$ can be expressed in terms of the cosmological constant $\Lambda$ as $H = \sqrt{\Lambda/3}$.  For
\begin{equation}
\label{Hcrit}
0<MH<MH_{\rm crit}\equiv 1/(3\sqrt{3})
\end{equation} 
there are two distinct horizons, the black hole horizon $R_{\rm bh}$ at
\begin{eqnarray} \label{Rbh}
\frac{R_{\rm bh}}{M} &=& -\frac{2}{\sqrt{3}MH}\cos\left(\frac{1}{3}\arccos\left(3\sqrt{3}MH\right)-\frac{2\pi}{3}\right) \nonumber\\&\approx& 2 + 8\left(MH\right)^{2} + \mathcal{O}\left(\left(MH\right)^{4}\right),
\end{eqnarray}
and the cosmological horizon $R_{\rm cos} > R_{\rm bh}$ at 
\begin{eqnarray} \label{Rc}
\frac{R_{\rm cos}}{M} &=& -\frac{2}{\sqrt{3}MH}\cos\left(\frac{1}{3}\arccos\left(3\sqrt{3}MH\right)+\frac{2\pi}{3}\right) \nonumber\\&\approx& \frac{1}{MH} - 1 +\mathcal{O}\left(MH\right)
\end{eqnarray}
(see Fig.~\ref{Rplots} below).

\subsection{McVittie coordinates}
\label{sec:mcvittie}

Alternatively, the line element (\ref{static}) can be expressed in McVittie coordinates \cite{McV33} as
\begin{equation} \label{mcvittie}
ds^{2} = -\left(\frac{1-\xi}{1+\xi}\right)^{2}dt^{2}_{{\rm M}} + a(t_{{\rm M}})^{2}\left(1+\xi\right)^{4}\left(d\bar{r}^{2}+\bar{r}^{2}d\Omega^{2}\right),
\end{equation} 
where $\xi\equiv M/\left(2a(t_{{\rm M}})\bar{r}\right)$ and $a(t_{{\rm M}})\equiv e^{Ht_{{\rm M}}}$, and where $\bar{r}$ is an isotropic radius.  In the limit of $H \rightarrow 0$ we recover the Schwarzschild metric in isotropic coordinates on slices of constant Schwarzschild time $T$.  We similarly recover the line element for a flat Friedmann-Lema\^itre-Robertson-Walker (FLRW) spacetime \cite{Fri22,Lem27,Rob35,Rob36a,Rob36b,Wal37} 
\begin{equation} \label{flrw}
ds^{2} = - dt^{2}_{{\rm M}} +  a(t_{{\rm M}})^{2}\left(d\bar{r}^{2}+\bar{r}^{2}d\Omega^{2}\right)
\end{equation}
in the limit $\bar{r}\rightarrow\infty$.   In the limit $\bar{r} \rightarrow 0$, the metric (\ref{mcvittie}) takes the same form as (\ref{flrw}),
just with a different scale factor; we therefore identify the McVittie coordinates (\ref{mcvittie}) with a ``wormhole" slicing of SdS spacetimes.

\subsection{Constant mean curvature slices}
\label{CMCSlices}

We now adopt a height-function approach to derive a family of time-independent constant mean curvature slices of SdS spacetimes (see also \cite{NakMNO91,BeiH05}, where this family was previously derived).   We will specialize to trumpet slices in Section \ref{CMCTrumpets}.

We begin with the static coordinates (\ref{static}) and transform to a new time coordinate $t$ with the help of a height function $h(R)$,
\begin{equation}
T = t- h(R).
\end{equation}
Only the derivative $h'\equiv dh(R)/dR$ will appear in the equations to follow.  Since $dT = dt-h'dR$ 
the new line element is
\begin{equation}
ds^{2} = -f dt^{2}+2h'fdtdR+\left(f^{-1}-fh'^{2}\right)dR^{2}+R^{2}d\Omega^{2}.
\end{equation}
Comparing terms with metric in the ``3+1" form
\begin{equation}
ds^{2} = \left(-\alpha^{2}+\beta_{i}\beta^{i}\right)dt^{2}+2\beta_{i}dx^{i}dt+\gamma_{ij}dx^{i}dx^{j},
\end{equation}
we can identify the lapse function $\alpha$, the shift vector $\beta^i$, and the spatial metric $\gamma_{ij}$.  In particular, we first note that $\beta_{R}=h'f$ and identify the metric components $\gamma_{RR}=f^{-1}-fh'^{2}$, $\gamma_{\theta\theta}=R^{2}$, and $\gamma_{\phi\phi}=R^{2}\sin^{2}\theta$.  We then invert the metric and raise the index of the shift vector to find $\beta^{i}=\left(\beta^{R},0,0\right)$ with $\beta^{R}=\gamma^{RR}\beta_{R}=\left(f^{-1}-fh'^{2}\right)^{-1}h'f$.  Finally we find the lapse from
\begin{equation}
-f = -\alpha^{2}+\beta_{R}\beta^{R} = -\alpha^{2}+h'f\left(f^{-1}-fh'^{2}\right)^{-1}h'f,
\end{equation}
so that
\begin{equation} \label{lapse}
\alpha^{2} = f+\frac{h'^{2}f^{2}}{f^{-1}-fh'^{2}} = \frac{1}{f^{-1}-fh'^{2}} = \frac{1}{\gamma_{RR}} = \gamma^{RR}.
\end{equation}
The extrinsic curvature $K_{ij}$ of the new slices is defined as $K_{ij} \equiv \gamma_i{}^a \gamma_j{}^b \nabla_a n_b$, where $\nabla_a$ is the covariant derivative associated with the spacetime metric $g_{ab}$, and where $n^{a}=\left(1,-\beta^{i}\right) / \alpha$ is the normal vector on the slices.   The mean curvature, defined as the trace of the extrinsic curvature, can then be computed from 
\begin{eqnarray}
K &=& -\nabla_{a}n^{a}=-|g|^{-1/2}\partial_{a}\left(|g|^{1/2}n^{a}\right)\nonumber\\&=&|g|^{-1/2}\partial_{R}\left(|g|^{1/2}\beta^{R}/\alpha\right), \label{defK}
\end{eqnarray}
We now fix $h'$ by requiring that slices of constant $t$ have the same mean curvature as slices of constant $t_{M}$ of the FLRW line element (\ref{flrw}), namely 
\begin{equation} \label{meancurvature}
K=K_{0}=-3H.
\end{equation}
Inserting this condition into (\ref{defK}) yields
\begin{equation}
-3HR^{2}=\partial_{R}\left(R^{2}h'f/\sqrt{f^{-1}-fh'^{2}}\right),
\end{equation}
which can be integrated to yield
\begin{equation}
-HR^{3}+C=\frac{R^{2}h'f}{\sqrt{f^{-1}-fh'^{2}}}
\end{equation}
where $C$ is a constant of integration.  Solving for $h'$ we obtain
\begin{equation}
h'^{2}f^{2}=\frac{\left(-HR^{3}+C\right)^{2}}{fR^{4}+(-HR^{3}+C)^{2}},
\end{equation}
and inserting this into (\ref{lapse}) results in
\begin{equation}
\alpha^{2} = \frac{f}{1-f^{2}h'^{2}}=\frac{fR^{4}+\left(-HR^{3}+C\right)^{2}}{R^{4}}.
\end{equation}
or
\begin{equation}
\label{lapse2}
\alpha = \sqrt{1-\frac{2M_{\rm eff}}{R}+\frac{C^{2}}{R^{4}}}.
\end{equation}
Here we have chosen the positive sign, and we have defined 
\begin{equation} \label{Meff}
M_{\rm eff}\equiv M+CH
\end{equation}
as a notational convenience.  Using $\beta^{R}=h'f\alpha^{2}$, we see that 
\begin{equation}
\beta^{R} = \pm\left(\frac{C\alpha}{R^{2}}-H\alpha R\right).
\end{equation}
Note that for large $R$ the magnitude of $\beta^R$ increases linearly with $R$ when $H > 0$.

Because the spatial metric remains time independent, the extrinsic curvature can be computed from
\begin{equation}
K_{ij}=\frac{1}{2\alpha}\left(D_{i}\beta_{j}+D_{j}\beta_{i}\right),
\end{equation}
where $D_i$ is the covariant derivative associated with the spatial metric $\gamma_{ij}$.  We find
\begin{equation}
K_{RR}=-\frac{2C+HR^{3}}{\alpha^{2}R^{3}}, ~~~ K_{\theta\theta} = \frac{K_{\phi\phi}}{\sin^{2}\theta}=\frac{C-HR^{3}}{R}.
\end{equation}
It is straightforward to verify that the mean curvature $K = \gamma^{ij}K_{ij}$ is indeed given by (\ref{meancurvature}).  


\subsection{Constant mean curvature trumpets}
\label{CMCTrumpets}

\begin{figure}
\centering
\includegraphics[width=3.3in]{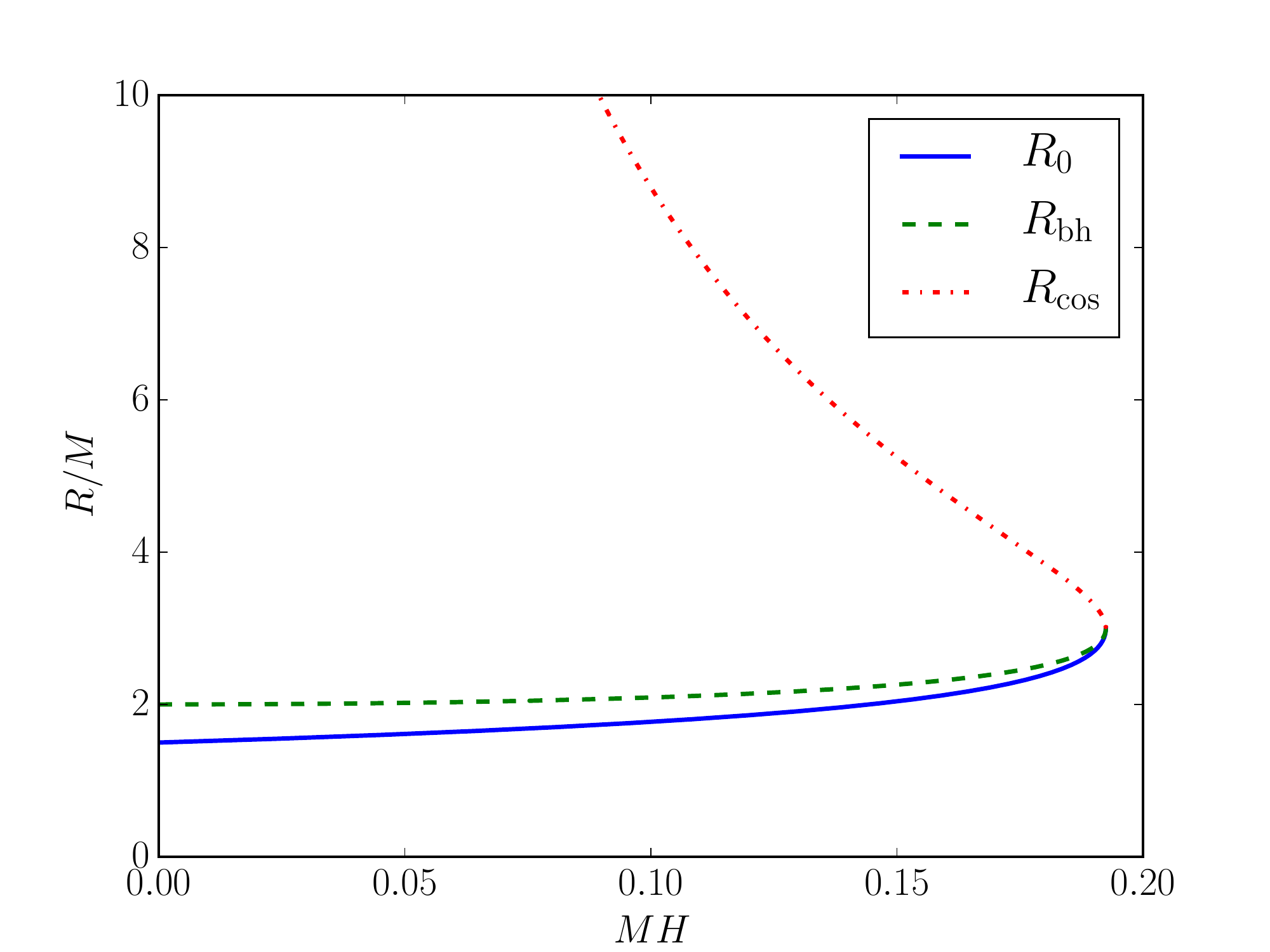}
\caption{The areal radius of the cosmological horizon $R_{\rm cos}$ (given by Eq.~(\ref{Rc}); dash-dotted line, red online), the black hole horizon $R_{\rm bh}$ (Eq.~(\ref{Rbh}); dashed line, green online), and the trumpet slices' limiting surface $R_0$ (Eq.~(\ref{R0}); solid line, blue online) as functions of the Hubble parameter $H$.}
\label{Rplots}
\end{figure}

The constant mean curvature slicing derived in Section \ref{CMCSlices} generalizes a family of maximal slices of the Schwarzschild spacetime, parametrized by $C$.  The special member of this family describing a trumpet slicing can be identified by requiring that the function $\alpha^2$ have a double root at some $0 < R_0 < R_{\rm bh}$ (see \cite{HanHPBO06,HanHOBO08,DenBM14}).  Applying the same criterion here, we write the square of  (\ref{lapse2}) as 
\begin{equation}
R^{4}-2M_{\rm eff}R^{3}+C^{2}=\left(R-R_{0}\right)^{2}\left(A_{2}R^{2}+A_{1}R+A_{0}\right),
\end{equation}
where we have introduced three new unknown constants $A_2$, $A_1$ and $A_0$.
Matching coefficients for the five different powers of $R$, we find that $A_{2}=1$, $A_{1}=M_{\rm eff}$, $A_{0}=3M_{\rm eff}^{2}/4$, as well as 
\begin{equation} \label{R0}
R_{0}=\frac{3M_{\rm eff}}{2}
\end{equation}
and 
\begin{equation} \label{C}
C^{2}=\frac{27M_{\rm eff}^{4}}{16}.
\end{equation}  
These expressions take the exact same form as in the asymptotically flat case, when $H = 0$ and  $M_{\rm eff} = M$.
Since, from its definition (\ref{Meff}), $M_{\rm eff}$ is a linear function of $C$, Eq.~(\ref{C}) is a quartic equation for $C$, but fortunately it is one that is easy to solve by hand.  There are four distinct solutions; two diverge as $H\rightarrow 0$, one has $C=-3\sqrt{3}M^{2}/4$ in that limit, and the other has the expected limit of $C=3\sqrt{3}M^{2}/4$.  Choosing that last solution we have
\begin{equation}
C = -\frac{M}{H}+\frac{2}{3\sqrt{3}H^{2}}\left(1-\sqrt{1-3\sqrt{3}MH}\right).
\end{equation}
Inserting this value into (\ref{Meff}) and (\ref{R0}) then yields the areal radius $R_0$ of the trumpet slices' limiting surface.  As shown in Fig.~\ref{Rplots}, we have $ 0 < R_{0}<R_{{\rm bh}}<R_{{\rm c}}$ for $0<H<H_{\rm crit}$, as desired.

\subsection{Isotropic coordinates}
\label{sec:iso}

For numerical purposes it is convenient to introduce an isotropic radius $r$.  We start by writing the spatial line element both in terms of the areal radius $R$ and the isotropic radius $r$, 
\begin{equation} \label{transform}
dl^{2}=\alpha^{-2}dR^{2}+R^{2}d\Omega^{2}=\psi^{4}_{{\rm static}}\left(dr^{2}+r^{2}d\Omega^{2}\right),
\end{equation}
where $\gamma_{RR} = \alpha^{-2}$ from Eq.~(\ref{lapse}).   Comparing the angular parts of (\ref{transform}) we find
\begin{equation} \label{psiRr}
\psi^{2}_{{\rm static}} = \frac{R}{r},
\end{equation}
and using this in a comparison of the radial parts of (\ref{transform}) yields
\begin{equation} \label{droverr}
\frac{RdR}{\sqrt{R^{4}-2M_{\rm eff}R^{3}+C^{2}}}=\pm\frac{dr}{r}.
\end{equation}
The isotropic coordinate takes the same form as it does for the maximal Schwarzschild trumpet \cite{BauN07}, just for different values for the mass and $C$.  Choosing the positive sign in Eq.~(\ref{droverr}) and picking the integration constant so that $r\rightarrow R$ as $R\rightarrow\infty$ we find
\begin{equation}
\label{rofR}
r = \frac{2R+M_{\rm eff}+\Xi}{4}\left(\frac{\left(4+3\sqrt{2}\right)\left(2R-3M_{\rm eff}\right)}{8R+6M_{\rm eff}+3\sqrt{2}\;\Xi}\right)^{1/\sqrt{2}},
\end{equation}
where we have defined $\Xi\equiv (4R^{2}+4M_{\rm eff}R+3M_{\rm eff}^{2})^{1/2}$.  Inserting this into (\ref{psiRr}) we obtain
\begin{eqnarray}
\psi_{{\rm static}} &=& \left(\frac{4R}{2R+M_{\rm eff}+\Xi}\right)^{1/2}\times\nonumber\\&&\left(\frac{8R+6M_{\rm eff}+3\sqrt{2}\;\Xi}{\left(4+3\sqrt{2}\right)\left(2R-3M_{\rm eff}\right)}\right)^{1/2\sqrt{2}}.
\end{eqnarray}
The origin of the isotropic coordinates, $r = 0$, corresponds to the limiting surface at (\ref{R0}), and it can be shown that the conformal factor $\psi_{{\rm static}}$ diverges there with $r^{-1/2}$.  This, in turn, guarantees that any point at $r > 0$ is an infinite proper distance removed from the origin at $r = 0$.  At large separations from the black hole, $\psi_{\rm static}$ approaches unity.

In these new coordinates, the lapse is still given by equation (\ref{lapse2}), with $R$ now understood as $R(r)$, which can be obtained by inverting equation (\ref{rofR}) numerically.  The radial component of the shift is
\begin{equation}
\beta^{r}=\beta^{R}\frac{dr}{dR}=\pm\frac{C-HR^{3}}{\psi^{2}_{{\rm static}}R^{2}},
\end{equation}
and the nonzero components of the extrinsic curvature are given by
\begin{equation}
K_{rr} = K_{RR}\left(\frac{dR}{dr}\right)^{2} = -\frac{2C+HR^{3}}{Rr^{2}},
\end{equation}
and
\begin{equation}
\label{extcurvangularcomp}
K_{\theta\theta} = \frac{K_{\phi\phi}}{\sin^{2}\theta}=\frac{C-HR^{3}}{R}.
\end{equation}
Expressed in these coordinates the solution is still time-independent.  However, at large distances from the black hole, the magnitude of the shift again increases linearly with $r$ when $H>0$, so that this solution is not compatible with the fall-off boundary conditions that are often imposed in numerical relativity. 

\subsection{McVittie-trumpet coordinates}
\label{sec:mcvittietrumpet}

Finally, in order to cast our trumpet solution in a McVittie-like form, we abandon our static trumpet coordinates and switch to a new ``comoving" isotropic radial coordinate $\bar{r}$ defined by
\begin{equation}
r=a(t)\, \bar{r}.
\end{equation}
Here $a(t)$ is given by $e^{Ht}$.
Note that this changes the coordinates within each slice, but does not change the slicing itself.  In terms of this new coordinate, the line element becomes
\begin{equation}
ds^{2} = -\left(1-\frac{2M_{\rm eff}}{R}\right)dt^{2}+\frac{2C}{R\bar{r}}dtd\bar{r}+\frac{R^{2}}{\bar{r}^{2}}\left(d\bar{r}^{2}+\bar{r}^{2}d\Omega^{2}\right),
\end{equation}
where $R$ is understood to be given as a function of $r = a(t) \,\bar r$.  The lapse is still given by equation (\ref{lapse}), while the radial component of the shift becomes
\begin{equation} \label{shift_mcvittie}
\beta^{\bar{r}}=\gamma^{\bar{r}\bar{r}}\beta_{\bar{r}}=\frac{\bar{r}^{2}}{R^{2}}\frac{C}{R\bar{r}}=\frac{C\bar{r}}{R^{3}}.
\end{equation}
Unlike in the previous coordinate systems, the shift vector now drops to zero for large $\bar r$, so that it is compatible with the fall-off boundary conditions often used in numerical relativity.  We also identify a new conformal factor $\psi$ using
\begin{equation}
\psi^{4}=\frac{R^{2}}{\bar{r}^{2}}=a^{2}\psi^{4}_{{\rm static}}.
\end{equation}
The nonzero components of the extrinsic curvature are given by
\begin{equation}
K_{\bar{r}\bar{r}}=-\frac{2C+HR^{3}}{\bar{r}^{2}R}
\end{equation}
together with Eq.~(\ref{extcurvangularcomp}).  We can again verify that the mean curvature $K$ is still given by (\ref{meancurvature}).

\begin{figure}
\centering
\includegraphics[width=3.3in]{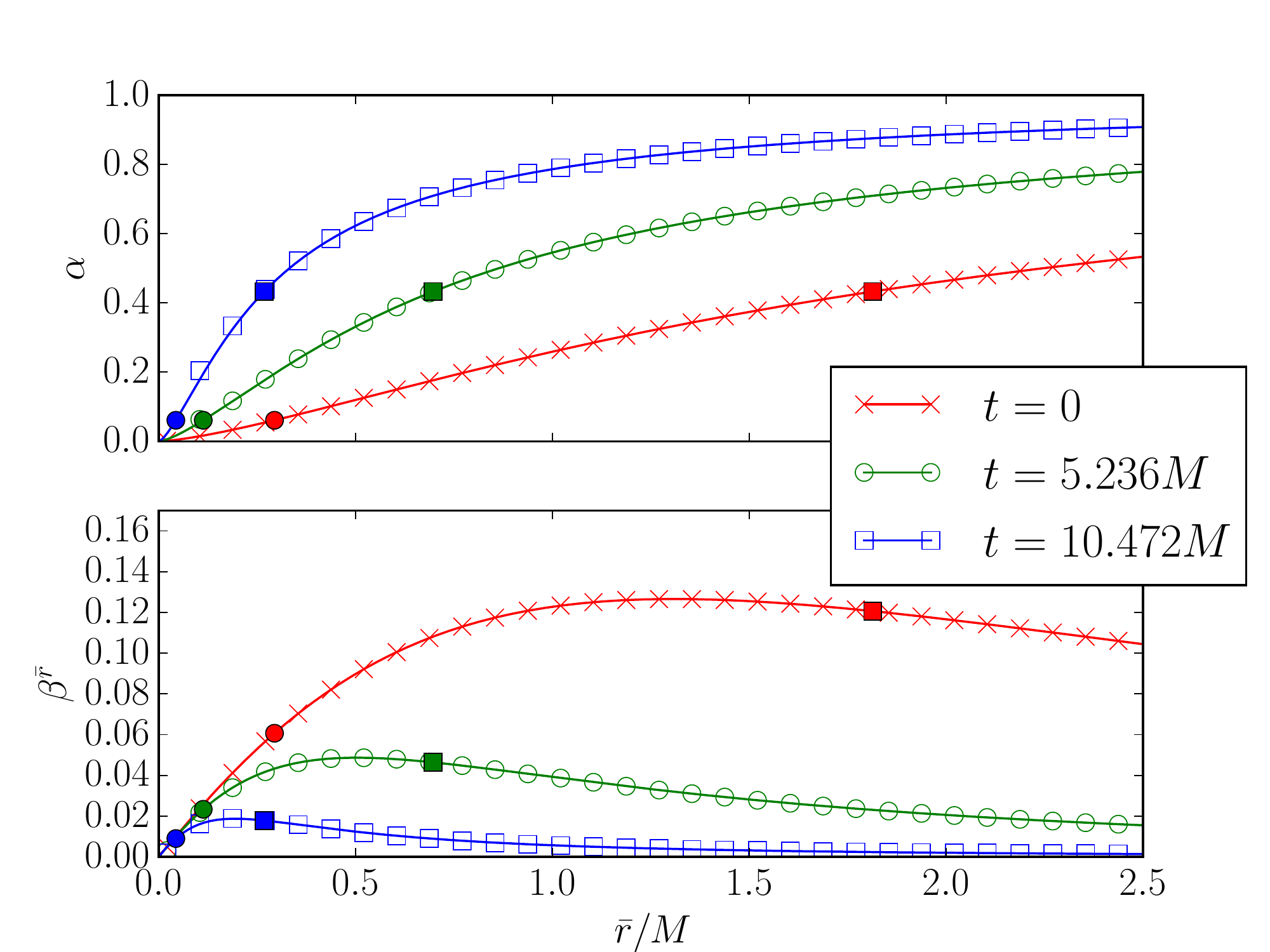}
\caption{The lapse (top panel) and radial component of the shift vector (bottom panel) at $t=0$, $t=5.236M$, and $t=10.472M$ for $H = 0.183 M^{-1} = \sqrt{0.9}H_{\rm crit}$ (see Eq.~(\ref{Hcrit})).  The analytical expressions are plotted as solid lines (red, green, and blue online).  The locations of the black hole and cosmological horizons are marked with filled circles and filled squares, respectively.  As a consequence of the large value of $H$ chosen here the two horizons are quite close to each other (compare Fig.~\ref{Rplots}).  The numerical results (crosses, open circles, and open squares) are from dynamical simulations of the McVittie-trumpet data of Sec.~\ref{sec:mcvittietrumpet}, evolved with the coordinate conditions (\ref{slicing}) and (\ref{gauge}), using 1,536 radial gridpoints (the lowest resolution used in Fig.~\ref{convergence}) and an outer boundary at $\bar{r}=64M$.  For clarity, only every second gridpoint is shown.  Agreement with the analytical expressions is excellent (see also the convergence test in Fig.~\ref{convergence}).}
\label{vstime}
\end{figure}

The above coordinates are McVittie-like in the sense that we recover the flat FLRW line element (\ref{flrw}) for large $\bar r$.  The shift vector (\ref{shift_mcvittie}) also drops off to zero for large separations from the black hole.  Unlike the McVittie coordinates of Section \ref{sec:mcvittie}, however, our new coordinates take a trumpet form instead of a ``wormhole'' form as $\bar{r}\rightarrow 0$; in particular, the conformal factor diverges with $\bar{r}^{-1/2}$.  We believe that these properties make these new coordinates well-suited for numerical simulations.   However, in comparison to the static isotropic coordinates of Section \ref{sec:iso} the new McVittie-like coordinates also have a disadvantage, since they are time-dependent.   We demonstrate this time-dependence in Fig.~\ref{vstime}, where we show profiles of the lapse function $\alpha$ and the shift vector $\beta^{\bar r}$ at different instants of time for $H=\sqrt{0.9}H_{{\rm crit}}$ (see Eq.~(\ref{Hcrit})).

\section{Numerical Demonstration}
\label{sec:numerics}

As a brief numerical demonstration we evolve the above data in a fully dynamical simulation.  We use the numerical code described in \cite{BauMCM13,BauMM15}, which evolves Einstein's equation 
in the Baumgarte-Shapiro-Shibata-Nakamura formulation \cite{NakOK87,ShiN95,BauS98} in spherical polar coordinates.  All spatial derivatives are evaluated using fourth-order finite-difference stencils (with the exception of advective terms, which are implemented to third order), but time derivatives are evaluated using a scheme that is accurate to second order only.   We follow \cite{ShiS99} in implementing an asymptotically dS spacetime; specifically, we write the conformal factor as $\psi^4 = a^2(t) \exp(4 \phi)$ with $a(t) = \exp(H t)$.  At large radii, where we impose the boundary condition $\phi \rightarrow 0$, our solutions then approach the FLRW line element (\ref{flrw}).

One complication arises from the fact that trumpet slices expressed in McVittie-like coordinates are time-dependent, which has to be taken into account when imposing coordinate conditions for the lapse and shift.  

In asymptotically flat spacetimes, dynamical simulations of black holes settle down to maximally sliced trumpets when the lapse function is evolved with a ``non-advective" version of the 1+log slicing condition $\partial_t \alpha = - 2 \alpha K$ (see \cite{BonMSS95,HanHOBGS06,BauN07}).  As suggested by \cite{ZilCGHSW12}, it is natural to replace this with $\partial_t \alpha = - 2 \alpha (K - K_0)$ for simulations in asymptotically dS spacetimes.  As $K \rightarrow K_0$, this results in $\partial_t \alpha \rightarrow 0$.  Unfortunately this condition cannot result in the McVittie trumpet slices of Sec.~\ref{sec:mcvittietrumpet}, since, for the latter, the lapse is {\em not} independent of time.  In fact, we find that
\begin{equation} \label{alpha_dot}
\partial_{t}\alpha = \frac{H}{2}\left(1-\alpha^{2}-3 \psi^4 \left( \beta^{\bar r }\right)^{2} \right),
\end{equation}
where we have used expressions for the lapse, shift, and conformal factor to eliminate $M_{\rm eff}$, $C$, and $R$.

A slicing condition that (i) reduces to a 1+log condition in the limit $H \rightarrow 0$ and (ii) captures the time-dependence (\ref{alpha_dot}) when $K = K_0$ is therefore given by
\begin{equation} \label{slicing}
\partial_{t}\alpha = \frac{H}{2}\left(1-\alpha^{2}-3 \psi^4 \left(\beta^{\bar{r}}\right)^{2}\right)-2\alpha\left(K-K_{0}\right).
\end{equation}
We impose this slicing condition in our dynamical simulations.  

From the expressions in Sec.~\ref{sec:mcvittietrumpet} we similarly find
\begin{equation}
\partial_{t}\beta^{\bar{r}}=\alpha\beta^{\bar{r}}K_{0}.
\end{equation}
This suggests the shift condition
\begin{equation} \label{gauge}
\partial_{t}B^{\bar{r}} = \frac{3}{4}\partial_{t}\bar{\Gamma}^{\bar{r}}-\eta\beta^{\bar{r}}, ~~~ \partial_{t}\beta^{\bar{r}}=B^{\bar{r}}+\alpha\beta^{\bar{r}}K_{0},
\end{equation}
which reduces to a non-advective gamma-driver shift condition \cite{CamLMZ06} in the limit that $H\rightarrow 0$.

\begin{figure}
\centering
\includegraphics[width=3.3in]{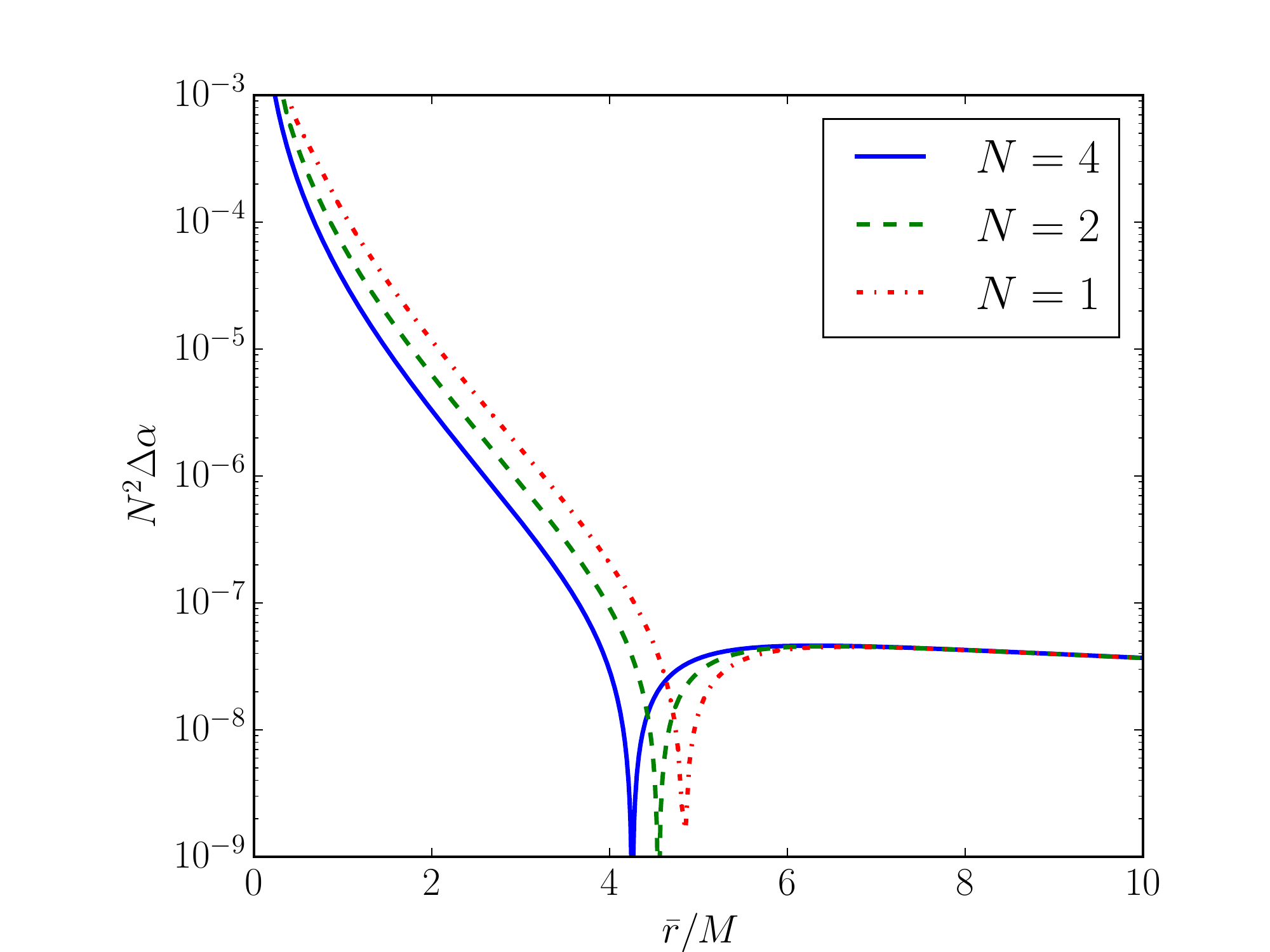}
\caption{The rescaled errors in the lapse (abs(analytical-numerical))  versus $\bar{r}/M$ at time $t=10.472M$ for $H=\sqrt{0.9}H_{{\rm crit}}$ (see Eq.~(\ref{Hcrit})).  We show results for three different numerical resolutions using 1,536$N$ radial gridpoints with $N = 1$, 2, or 4 (dash-dotted, dashed, and solid lines -- red, green, and blue online).  
The numerical errors decrease as expected for a second-order convergent code or slightly faster.}
\label{convergence}
\end{figure}

In Figs.~\ref{vstime} and \ref{convergence} we show results from dynamical simulations of the McVittie-trumpet data of Sec.~\ref{sec:mcvittietrumpet}, evolved with the coordinate conditions (\ref{slicing}) and (\ref{gauge}).  In Fig.~\ref{vstime} we show snapshots of the lapse function $\alpha$ and the radial component of the shift vector $\beta^{\bar r}$ at different instants of time for $H=\sqrt{0.9}H_{{\rm crit}}$ (\ref{Hcrit}), comparing the analytical expressions (lines) with numerical results (points).   In Fig.~\ref{convergence} we show differences between the analytical and numerical data for three different numerical resolutions, showing that the numerical errors decrease at least as fast as expected for a second-order convergent code.

We note that the coordinate conditions (\ref{slicing}) and (\ref{gauge}) are ``tailored" to the McVittie-trumpets of Sec.~\ref{sec:mcvittietrumpet}, and may not be suitable for other data.  In fact, we have found that dynamical evolutions of the McVittie-wormholes (\ref{mcvittie}) with these coordinate conditions will lead to numerical instabilities.

\section{Summary}
\label{Sum}

Motivated by numerical simulations of black holes in asymptotically dS spacetimes \cite{ZilCGHSW12} we study trumpet slices in SdS spacetimes.  Starting with the line element for SdS spacetimes in static coordinates we follow a sequence of coordinate transformations and identify special constant mean-curvature slices which we then express in isotropic McVittie-like coordinates \cite{McV33}.  These slices are natural generalizations of the maximal trumpet slices of the Schwarzschild spacetime \cite{HanHOBGS06,BauN07}: they cast the black hole in a trumpet geometry, while far away from the black hole the metric approaches the FLRW form (\ref{flrw}).   Unlike the trumpet slices of the Schwarzschild spacetime, however, McVittie-trumpets of SdS spacetimes do depend on time.   This time-dependence needs to be taken into account in coordinate conditions designed to reproduce these trumpets in dynamical simulations.  We perform such simulations, and demonstrate how our results can be used as a test for black hole simulations in asymptotically dS spacetimes.  

\acknowledgments

This work was supported in part by NSF grants 1402780 and 1707526 to Bowdoin College.


\end{document}